\documentclass[aps,prl,twocolumn,superscriptaddress,nolongbibliography]{revtex4-2}

\usepackage{graphicx} 

\usepackage{amsmath}
\usepackage{amssymb}
\usepackage{enumitem}
\usepackage{multirow}
\usepackage{physics}
\usepackage[dvipsnames]{xcolor}
\usepackage[T1]{fontenc}
\usepackage{graphicx}
\usepackage{dcolumn}
\usepackage{bm}
\usepackage[retainorgcmds]{IEEEtrantools}
\usepackage{xcolor}
\usepackage[colorlinks=true,linkcolor=blue,citecolor=blue,urlcolor=blue]{hyperref}
\usepackage{ulem}
\usepackage{bbold}
\usepackage{mathtools}
\usepackage{accents}
\usepackage{verbatim} 
\usepackage{dutchcal}
\usepackage{gensymb}

\newcommand{\chibar}{\overline{\chi}}

\newcommand{\jind}{\mathbf{j}^{\mathrm{ind}}}

\newcommand{\G}{\mathbf{G}}
\newcommand{\qz}{q_z}

\newcommand{\ZD}{\mathrm{2D}}

\newcommand{\vcoul}{v}
\newcommand{\qext}{\mathbf{q}}

\newcommand{\w}{\omega}
\renewcommand{\k}{\mathbf{k}}

\begin{document}

%
\nocite{apsrev42Control}

\title{Unified description of exciton, phonon and plasmon dispersions in 2D materials from an optical conductivity approximation}

\author{Alberto Guandalini}
\email{alberto.guandalini@uniroma1.it}
\affiliation{Dipartimento di Fisica, Universit\`a di Roma La Sapienza, Piazzale Aldo Moro 5, I-00185 Roma, Italy}

\author{Andrea Ferretti}
\affiliation{Istituto Nanoscienze – CNR, S3, Via G. Campi 213/A, Modena, Italy}

\author{Daniele Varsano}
\affiliation{Istituto Nanoscienze – CNR, S3, Via G. Campi 213/A, Modena, Italy}

\author{Paolo Barone}
\affiliation{Dipartimento di Fisica, Universit\`a di Roma La Sapienza, Piazzale Aldo Moro 5, I-00185 Roma, Italy}
\affiliation{CNR-SPIN, Area della Ricerca di Tor Vergata, Via del Fosso del Cavaliere 100, I-00133 Rome, Italy}

\author{Francesco Mauri}
\affiliation{Dipartimento di Fisica, Universit\`a di Roma La Sapienza, Piazzale Aldo Moro 5, I-00185 Roma, Italy}

\date{\today}

\begin{abstract}
The modified Coulomb interaction in two-dimensional (2D) materials gives rise to unique, non-analytic low-momentum dispersions of excitons, phonons, and plasmons. Here, we describe the 2D dispersion of longitudinal excitations by introducing the optical conductivity approximation (OCA), a unified framework that evaluates the finite-$\mathbf{q}$ longitudinal response probed by electron energy-loss spectroscopy (q-EELS) by only using the $\mathbf{q}=0$ optical conductivity $\sigma(\omega)$. 
By doing so, the approach demonstrates that the linear dispersion of all the above mentioned excitations is mainly due to the form of the 2D macroscopic Coulomb interaction rather then to the dispersion of the underlying band structure. Applying this approach to hexagonal boron nitride and graphene we accurately reproduce the energy and intensity dispersions of the hBN longitudinal-optical phonon and  low-energy bright excitons, along with the graphene $\pi$-plasmon. Our work also provides an efficient computational scheme to describe low momentum EELS data without the need to calculate the response functions at dense momentum grids.
\end{abstract}
\maketitle

Crystal elementary excitations, such as excitons, phonons, and plasmons -- and their momentum dispersions -- are key quantities in condensed matter physics.
Longitudinal excitations constitute the poles with non-vanishing oscillator strengths of the density-density response function $\chi$, that connects an external scalar perturbation to the induced charge (electron plus ionic) density of the system. 
Momentum-resolved electron energy-loss spectroscopy ($q$-EELS) 
is a powerful tool to study the dispersion of longitudinal excitations, as its scattering cross-section is proportional to the imaginary part of the momentum- and energy-resolved density-density response function Im$[-\chi(\mathbf{q},\omega)]$~\cite{Egerton_book,Ibach_book}. In this respect, recent advances in transmission electron microscope $q$-EELS provide unprecedented opportunities for exploring fundamental excitations in freestanding, truly two-dimensional (2D) systems~\cite{Senga_2019, guandalini_2023, Guandalini_mes_nothing, Jinhua_2025, Advances_qEELS_PRM2026}.

Since these excitations are longitudinal, they couple directly to the long-range macroscopic component of the (also longitudinal) Coulomb interaction $v$, as opposite to transverse excitations~\cite{Denisov_73,Baroni_01,Qiu_15,Sohier_15}.
In bulk three-dimensional (3D) crystals, this difference 
induces a finite energy splitting between longitudinal and transverse modes in the long-wavelength limit ($\mathbf{q}\to 0$), leading to the conventional longitudinal-transverse optical (LO/TO) splitting for phonons~\cite{Baroni_01,Royo_21}, and to an analogous finite splitting for excitons~\cite{qiu_21,Nalabothula_26}. In two-dimensional (2D) materials, this scenario changes radically because the reduced dimensionality modifies the long-range behavior of the Coulomb potential from 
$v^{\mathrm{3D}}(\mathbf{q}) \sim 4\pi/q^2$ to $ v^{\mathrm{2D}}(\mathbf{q}) \sim 2\pi/q$.
As a first consequence of this macroscopic field weakening,
in the optical limit ($\mathbf{q}\to 0$) both the direct and inverse dielectric functions approach unity~\cite{Cudazzo_2011,Huser_2013}, i.e. $\epsilon^{-1,\ZD}(\mathbf{q}\to 0,\omega) = \epsilon^{\ZD}(\mathbf{q}\to 0,\omega) = 1$. Therefore, the optical absorption spectrum and the $\mathbf{q}\approx 0$ EEL cross-section become proportional, aside from a trivial $\omega$ factor.

Due to this, the finite 3D splitting breaks down: the LO/TO phonon splitting collapses exactly at the $\Gamma$ point ($\mathbf{q}=0$)~\cite{Thibault_2017}, as also observed for bright (longitudinal) and dark (transverse) excitons~\cite{Qiu_15,qiu_21}.
Additionally, longitudinal phonons/excitons exhibit a non-analytic linear dispersion ($\propto q$) whereas the transverse ones retain a conventional parabolic dispersion ($\propto q^2$)~\cite{Qiu_15, qiu_21,liu2026direct, Thibault_2017}.
For the same dimensionality scaling of the Coulomb interaction, the dispersion of 2D Drude plasmons becomes acoustic~\cite{Hwang_07,daJornada_20}.
Finally, the static screening in 2D materials is strongly q-dependent in Fourier space~\cite{Thygesen_2017,Macheda_2024,Guandalini_2024}, 
often requiring the use of dedicated 
computational approaches to calculate the quasiparticle electronic structures~\cite{Rasmussen_2016,daJornada_2017,Xia_2020,Guandalini_2024}, excitons~\cite{Cudazzo_2011,qiu2016screening,Trolle_17}, and phonons~\cite{Sohier_15,Royo_21}, with respect to their 3D counterpart.
Despite the extensive effort to describe the above-mentioned 2D-specific phenomena  and their effect over electron and lattice excitations, a unified framework accounting for all these effects is still missing. 

In this work, we resolve this fragmentation by establishing a formal link between the finite-$\mathbf{q}$ longitudinal response and the $\mathbf{q}=0$ optical limit of 2D systems.
In particular, (momentum-resolved) $\mathbf{q}$-EELS strictly probes the longitudinal density-density response, which can be expressed via the momentum-dependent conductivity $\sigma(\mathbf{q},\omega)$. 
Being inherently free from the macroscopic (long-range) Coulomb potential in the long-wavelength limit ($\mathbf{q}\to 0$), one can expect the conductivity $\sigma(\mathbf{q},\omega)$ to display a milder $\mathbf{q}$-dependence than the reducible density-density response $\chi(\mathbf{q},\omega)$ or, equivalently, the inverse dielectric function $\epsilon^{-1}(\mathbf{q},\omega)$.
In view of this, we introduce the optical conductivity approximation (OCA)
to accurately describe the low-momentum finite-$\mathbf{q}$ density-density response $\chi$ of 2D systems using only the optical conductivity.

By applying OCA to monolayer and bilayer h-BN~\cite{Zhang_2017,wang2019epitaxial,chen2020wafer,caldwell2019photonics}, and graphene~\cite {Novoselov_2016,novoselov_two-dimensional_2005}, we show that the energy and intensity dispersions of LO phonons, bright excitons~\cite{Jinhua_2025}, and plasmons~\cite{guandalini_2023} are accurately reproduced. 
This demonstrates that their sharp low-momentum dispersions are not driven by intrinsic band-structure variations, being rather a pure consequence of the 2D macroscopic Coulomb interaction.
Notably, OCA can be applied to any pole of the density-density response, including satellites generated by interactions such as the exciton-phonon coupling~\cite{Jinhua_2025}, providing both an efficient computational scheme to bypass dense momentum grids and an inverse method to extract the complex optical conductivity from low-momentum EEL spectra. 
Additionally, it also underpins and further clarifies the W-average method, a convergence accelerator for the integration of the screened Coulomb interaction in the GW framework, recently developed by some of the Authors~\cite{Guandalini2023npjCM,Guandalini_2024,Sesti_2026}.

\textit{EEL cross section of a 2D material---}
We consider a high energy electron beam incoming perpendicular to a periodic 2D material, for simplicity taken with zero thickness.

The straightforward generalization to finite thickness is covered in the Supplemental Material, and used in the numerical results of this work.
Within scattering theory in the Born approximation~\cite{Sturm_1993,Nazarov_2015,Senga_2019,Guandalini_mes_nothing}, the (transmission) EEL differential cross-section is given by\footnote{All the equations are given in atomic units.}: 
\begin{multline}\label{eq:cross_section}
\frac{d^2S}{d\Omega d\omega} =
\frac{4A} 
{\pi}\frac{1}{(|\qext+\G|^2+\qz^2)^2}
\mathrm{Im} \left[-
\chi_{\G\G}(\qext,\omega)\right],  
\end{multline}
where $A$ is the area of the 2D unit cell. Here, $\mathbf{q}+\G$ is the momentum transferred in the (in-plane) periodic directions of the crystal, $\qext$ is within the first Brillouin zone and $\G$ a reciprocal lattice vector, $q_z$ the momentum lost in the incident direction (still present even if the material is 2D~\cite{Guandalini_mes_nothing}), and $\omega$ the energy loss.
As we are interested in low momentum transfers, we consider only the $\G=0$ case and $\chi_{00}(\qext,\w)\equiv \chi(\qext,\w)$.
We will implicitly consider that quantities are evaluated at $\G=0$ unless explicitly stated.
The zero-thickness model is a valid assumption when $q\ll 1/d$, where $d$ is the physical material thickness~\footnote{As $\chi$ is a density response function, it is the density extension of the system $d$ which matters. In the case of the extension of the induced potentials, the dielectric extension $r_{\mathrm{eff}}$ must be considered instead~\cite{Thibault_16}}.

The density-density response function $\chi(\qext,\omega)$ is the Fourier transform of the derivative of the induced density $\rho^{\mathrm{ind}}$, containing both the electron and lattice contributions
(thus including also the lattice response shown in EELS~\cite{Senga_2019}), with respect to the external scalar perturbation $v^{\mathrm{ext}}$,
$\chi(\qext,\omega) = \delta\rho^{\mathrm{ind}}(\qext,\omega)/\delta v^{\mathrm{ext}}(\qext,\omega)$
and has poles corresponding to different longitudinal excitations, such as plasmons, excitons and phonons. 

\textit{Optical Conductivity Approximation (OCA)---}
The dispersion of the $\chi(\qext,\w)$ poles arises from two main contributions.
The first is the dispersion of the underlying band structure, that directly affects electron-hole (e-h) excitations 
involving quasiparticle orbitals $\psi_{v\k}$ and $\psi_{c\k+\qext}$, for valence and conduction bands, and shapes both exciton~\cite{Gatti_13,Reining_16} and phonon responses~\cite{Macheda_2024_2, Caldarelli_2024, Guandalini_phBSE}. 
The second contribution is the momentum-dependence of the macroscopic Coulomb interaction $v^{\ZD}=2\pi/q$,  present in the electronic Hartree potential response.

While the density-density response $\chi(\mathbf{q},\omega)$ exhibits sharp variations near the $\Gamma$ point due to the long-range macroscopic Coulomb interaction, this singular q-dependence can be formally isolated by introducing the (barred) $\chibar$~\cite{Gatti_13, Macheda_2024_2} response as the variation 
with respect to the total macroscopic potential rather than the external one.

In order to do so, we first connect the density-density response $\chi(\mathbf{q},\omega)$ to the conductivity $\sigma_{\alpha\beta}(\qext,\omega) = \delta j_{\alpha}^{\mathrm{ind}}(\qext,\omega)/\delta E_{\beta}^{\mathrm{tot}}(\qext,\omega)$,
where $\alpha,\beta=(x,y)$ are Cartesian coordinates and 
$\sigma_{\alpha\beta}$ is the momentum-dependent conductivity, connecting the induced charge current $j^{\mathrm{ind}}_{\alpha}$(both electronic and ionic) with the total macroscopic electric field  $\mathbf{E}^{\mathrm{tot}}(\qext,\omega) = i\mathbf{q}v^{\mathrm{tot}}(\qext,\omega)$.
Local field ($\G \neq 0$) corrections are explicitly included in the conductivity, as described in the Supplemental Material.
We set $v^{\mathrm{tot}}(\qext,\omega) = v^{\mathrm{ext}}(\qext,\omega)+v^{\mathrm{ind}}(\qext,\omega)$, where $v^{\mathrm{ind}}$ is the Hartree potential generated by $\rho^{\mathrm{ind}}$.
Exploiting the continuity equation $-i\w\rho^{\text{ind}}(\qext,\w)+i\qext\cdot\jind(\qext,\w) = 0$,
a general expression relating $\chi$ and $\sigma$ can be drawn 
\begin{equation}\label{eq:OCAq}
    \chi(\qext,\w) = \frac{\sigma(\qext,\w)|\qext|^2}{-i\w-v^{\mathrm{2D}}(\mathbf{q}) \sigma(\qext,\w)|\mathbf{q}|^2} \ ,
\end{equation}
where $\sigma(\qext,\w) =\mathbf{q}\cdot\bm{\sigma}\cdot\mathbf{q}/|\mathbf{q}|^2$
is the longitudinal conductivity. 
Finite thickness effects can be described by approximating the profile of the electron density as rectangular~\cite{Macheda_23,TB_paper}, leading to an additional shape factor $F(|\mathbf{q}|)$ in the second term of the denominator of Eq.~\eqref{eq:OCAq}.
We refer to the Supplemental Material for an exhaustive derivation.
Importantly, Eq.~\eqref{eq:OCAq} highlights one of the main reasons to work with $\sigma$: 
In the optical limit $\sigma \to \mathrm{const.}$, while $\chi \to 0$ as $q^2$, making $\sigma$ an easier object to approximate as a function of momentum.
In addition, the momentum dependence of $\chi$ originating from the macroscopic Hartree term is analytically treated by the denominator. 

Furthermore, the continuity equation links $\chi(\qext,\w)$ directly to the conductivity $\sigma(\mathbf{q},\omega)$, explicitly factoring out the remaining kinematical $\mathbf{q}$-dependencies. As a result, $\sigma(\mathbf{q},\omega)$ acts as a response to the total macroscopic field and is inherently free from the macroscopic Coulomb divergence. Consequently, as previously noted in the literature~\cite{Nazarov_15} and confirmed by our results below, $\sigma(\mathbf{q},\omega)$ exhibits only a weak dependence on $\mathbf{q}$ near the $\Gamma$ point.

Guided by this physical behavior, we can safely approximate the finite-momentum conductivity with its optical counterpart (optical conductivity approximation, OCA) -- an assumption that would instead be fundamentally incorrect for $\chi$. 
Indeed, within OCA, we have $\sigma(\mathbf{q},\w) \approx \sigma(\mathbf{q}=0,\w)\equiv\sigma(\w)$ and 
\begin{equation}
\label{eq:OCA}
   \chi^{\mathrm{OCA}}(\qext,\w) = \frac{\sigma(\w)|\qext|^2}{-i\w-v^{\mathrm{2D}}(\mathbf{q})\sigma(\w)|\mathbf{q}|^2} \ . 
\end{equation}
By substituting Eq.~\eqref{eq:OCA} into Eq.~\eqref{eq:cross_section}  within  OCA, we find a direct relation between low-momenta EEL spectra and the optical conductivity.
This is the central results of this work.
Provided that OCA accurately describes the EELS cross-section, as validated later in this work, our results show that the dispersion of charged excitations in 2D materials is dominated by 
the momentum dependence of the Coulomb interaction rather than that of the optical conductivity.
We note that $\sigma(\omega)$ is in general a complex quantity, and the EEL spectra depend on both $\mathrm{Re}[\sigma]$ and $\mathrm{Im}[\sigma]$ through the denominator on the right side of Eq.~\eqref{eq:OCA}.
The OCA is exact at $\mathbf{q}=0$ and we show it to provide accurate results for a large range of momenta.

We decompose the conductivity into electronic (at fixed nuclei) and vibrational contributions, $\sigma = \sigma_{\mathrm{el}}+\sigma_{\mathrm{ph}}$.
The individual terms can be written as
\begin{equation}
    \sigma_{\mathrm{el}/\mathrm{ph}}(\w) = -i\w\left[\sum_{\lambda} \frac{R^{\mathrm{el}/\mathrm{ph}}_{\lambda}}{(\w+i\eta)^2-(\Omega^{\mathrm{el}/\mathrm{ph}}_\lambda)^2}+C\right],
\end{equation}
where $\eta$ is a small positive quantity, $\Omega^{\mathrm{el}/\mathrm{ph}}_{\lambda}$ are the energy of the electron/phonon resonances,  $R^{\mathrm{el/ph}}_{\lambda} = |\mu^{el/ph}_{\lambda}|^2/A$ are the corresponding residues with $\mu^{el/ph}_{\lambda}$ the dipole transition moment of excitation $\lambda$ and $A$ the area of the 2D unit cell.
For phonons, $\mu^{ph}_{\lambda}$ are given by the Born-effective charges~\cite{Macheda_2024_2}.
$C$ is a real constant, eventually taking into account higher-energy excitations not included in the $\lambda$ sum.
Within the proposed procedure, we are able to describe both electronic and ionic contributions. 

\textit{Results---}
In the following, we focus on EELS experiments and consider the loss functions $\mathrm{EEL(\mathbf{q},\omega)} \propto -\mathrm{Im[\chi(\mathbf{q},\omega)]}$ normalized at each $\mathbf{q}$, since these usually are the experimental output.
For the reference calculations, we split $\chi = \chi_{\mathrm{el}}+\chi_{\mathrm{ph}}$, where $\chi_{\mathrm{el}}$ is calculated with the Bethe-Salpeter Equation (BSE)~\cite{yambo_2009} and $\chi_{\mathrm{ph}}$ with the Density Functional Perturbation Theory (DFPT) framework~\cite{Giannozzi2009,Giannozzi2017,Senga_2019,Macheda_2024_2}. The optical conductivities $\sigma_{\mathrm{el}}/\sigma_{\mathrm{ph}}$, used in the OCA, have been obtained with the same methods.
Computational details can be found in the Supplemental Material.
In the following, we neglect kinetic contributions arising when $q\approx q_z$ by setting $q_z=0$ in Eq.~\eqref{eq:cross_section}, as these effects have been already studied in a previous work~\cite{Guandalini_mes_nothing}.
We focus on $\mathbf{q}$-vectors around the $\Gamma$ point,  where the graphene and hBN responses are nearly isotropic in the plane.

\begin{figure}[t] 
    \centering
    \includegraphics[width =\textwidth/2, keepaspectratio]{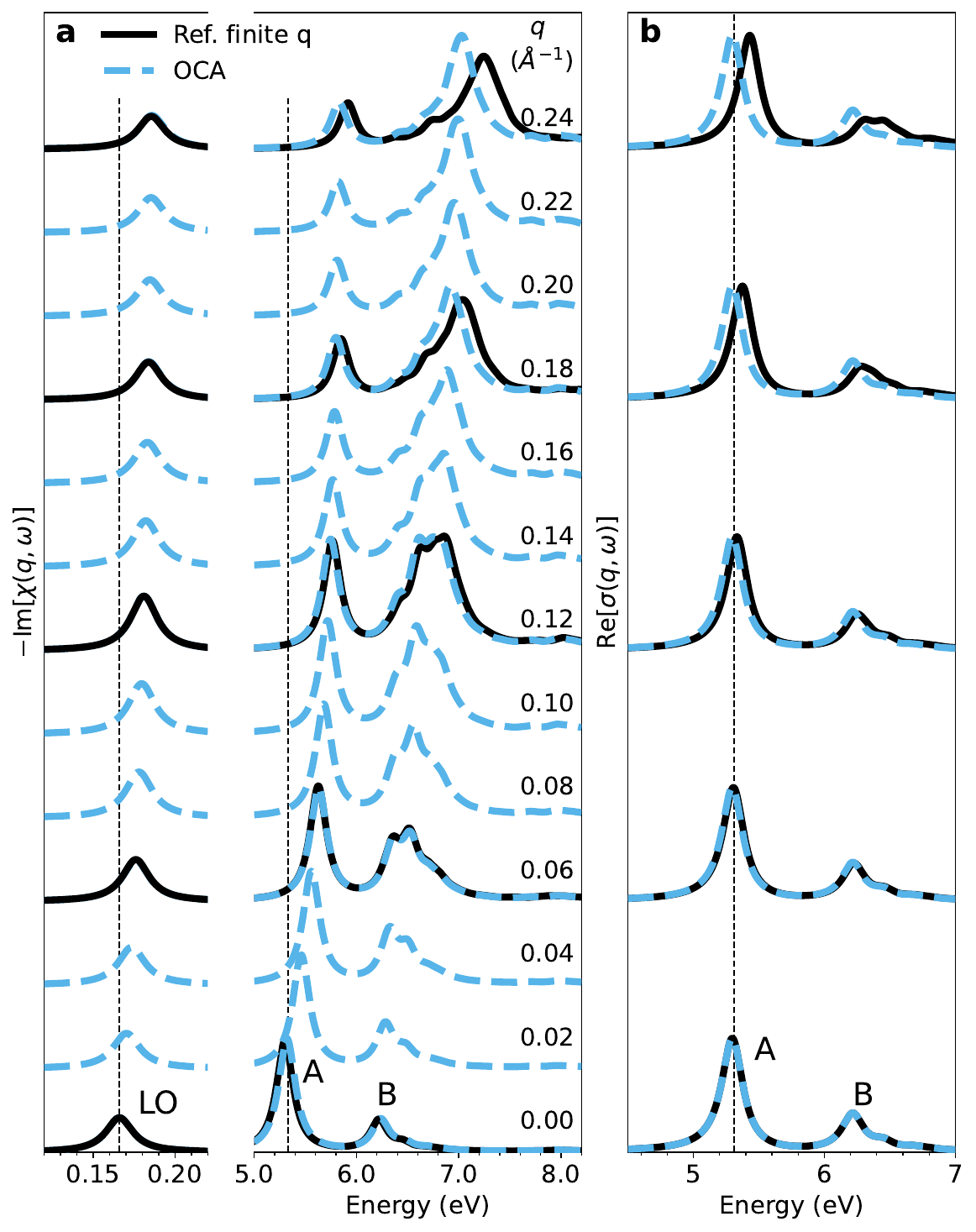}
    \caption{Ionic and electron loss functions (a) and electron conductivity (b) at different momentum transfers of monolayer hBN.
    Ab-initio calculations obtained at finite momentum transfer are in black and those obtained with the optical conductivity approximation (OCA) are in blue. Energy positions at $q=0$ are underlined with vertical light-blue dashed lines. Excitonic branches are labeled with letters (A and B) while the longitudinal-optical phonon excitation with LO. Ab-initio calculations are done along the $\Gamma$M direction.}
    \label{fig:spectra_hBN}
\end{figure}

We demonstrate the accuracy of OCA by showing in Fig.~\ref{fig:spectra_hBN}(a) the finite-$\mathbf{q}$ EEL spectra of monolayer hBN up to $q = 0.24$ \AA$^{-1}$ (for reference, $|\Gamma M|=1.45$ \AA$^{-1}$ and $|\Gamma K|=1.67$ \AA$^{-1}$).
In the low-energy part of the spectrum, we find a phonon excitation peak corresponding to the longitudinal-optical (LO) mode~\cite{Thibault_2017,Senga_2019}, 
accurately reproduced in terms of both intensity and dispersion by OCA. 
At higher energies, the EEL of monolayer hBN has a single-exciton peak ($A$) and a second peak ($B$) composed by several excitons with similar energies.
This picture is consistent with previous measurements~\cite{Jinhua_2025} and simulations~\cite{Paleari_2018}.
The same peak composition is observed for the electronic part of $\sigma(\mathbf{q},\omega)$ in Fig.~\ref{fig:spectra_hBN}(b).
The ionic part of the conductivity is not shown, as its dispersion is negligible in this momentum range, as evinced by the accuracy of OCA.

\begin{figure*}[t!] 
    \centering
    \includegraphics[width = 1.0\textwidth, keepaspectratio]{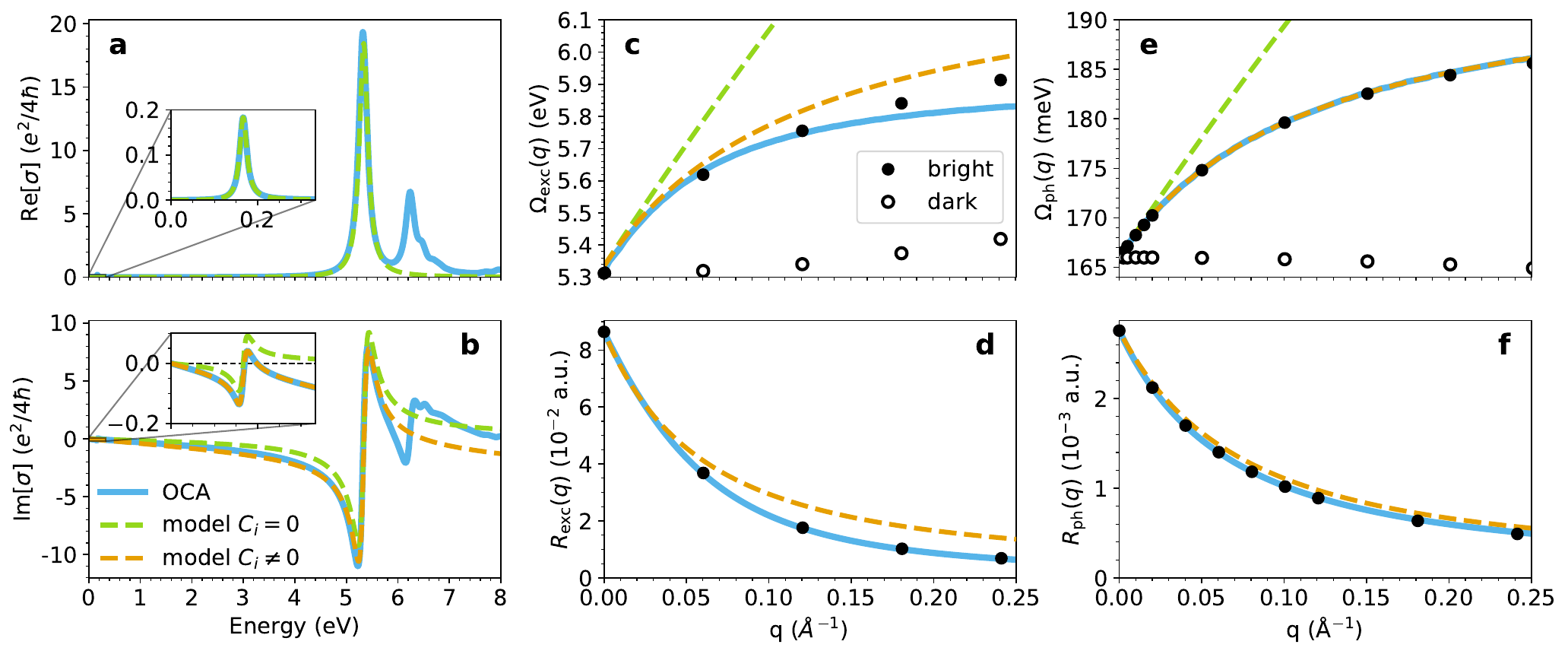}
    \caption{Real (a) and imaginary (b) parts of the optical conductivity of monolayer hBN. A exciton branch dispersion (c) and $\chi$ intensity of the bright A exciton (d).
    Optical phonon dispersions (e) and $\chi$ intensity (f) of the longitudinal optical (LO) phonon. We show in black the ab-initio calculations (bright/LO excitations with filled circles and dark/TO excitations with empty circles), while other results are obtained with the optical conductivity approximation (OCA).
    In blue, obtained with the full $\sigma(\omega)$ obtained from ab-initio calculations. In red (orange), results obtained with a model optical conductivity as in Eq.~\eqref{eq:sigma_mod} with $C_i=0$ ($C_i \neq 0$).}
    \label{fig:disp_hBN}
\end{figure*}

The EEL excitonic spectra show highly dispersive peaks, with a fast increase of the intensity ratio between B and A, as opposite to $\sigma(\mathbf{q},\omega)$, verifying a posteriori our ansatz about the OCA.
By construction, OCA is exact at $\mathbf{q}=0$, while its accuracy slightly deteriorates increasing momentum transfer.
Notably, OCA works better for the $A$ than the $B$ peak, due to the shape change of the $B$ peak in $\sigma(\qext,\omega)$, originated from its composite excitonic nature.
Remarkably, the change of shape of the $B$ peak with momentum transfer, is qualitatively reproduced by OCA.
From the OCA ability to reproduce all the finite-momentum features of the EEL spectra, we conclude the momentum dispersion of the conductivity does not play a role in the momentum dispersions of the exciton and phonon branches considered. Such dispersion properties are instead determined by the explicit momentum dependence given by Eq.~\eqref{eq:OCA}, originated by the continuity equation and the 2D dependence of the Coulomb interaction.

In Fig.~\ref{fig:disp_hBN}, we show the real (a) and imaginary (b) parts of the optical conductivity $\sigma(\omega) = \sigma_{\mathrm{el}}(\omega)+\sigma_{\mathrm{ph}}(\omega)$ of monolayer hBN along with a quantitative analysis of the energy and residue dispersions of the $A$ (c,d) excitons and LO/TO phonons (e,f).
We note the intensity ratio between the ionic  and electronic parts is much lower in the conductivity than in the EEL spectra, due to the $\omega$ factor between $\sigma$ and $\chi$.
There is a clear similar trend between the exciton and phonon dispersion/intensity behaviors.
Both quasi-particles are degenerate at $\mathbf{q}=0$, with a bright/longitudinal and a dark/transverse mode.
The bright mode linearly disperses up to a saturation, with quadratically-decaying intensity, while the dark one disperses at least quadratically. 
Such a different behavior depends on the quasi-particle interaction with the macroscopic Coulomb interaction, as already separately studied for excitons~\cite{Qiu_15} and phonons~\cite{Thibault_2017}.  
Importantly, OCA reproduces accurately both the $A$ exciton and the LO phonon dispersions, demonstrating that all the physics required to describe such processes is the optical conductivity $\sigma(\omega)$ and the $q^{-1}$ dependence of the 2D Coulomb interaction.
Of course, dark EELS excitations, such as the dark $A$ exciton and the TO phonon mode, cannot be described within the OCA framework.

To get further insight, we can approximate $\sigma$ near a single resonance $\Omega_{0,i}$ as a single Lorentzian plus a non-absorptive screening background corresponding to the higher-energy resonances as 

\begin{equation}\label{eq:sigma_mod}
    \sigma_{\mathrm{mod},i}(\w) = -i\w\left[\frac{R_{0,i}}{(\w+i\eta)^2-\Omega_{0,i}^2} + C_i\right] \ ,
\end{equation}
where $i=\mathrm{el},\mathrm{ph}$ and the parameters $R_{0,i}$, $\Omega_{0,i}$  are fitted on the ab-initio $\sigma$.
By substituting Eq.~\eqref{eq:sigma_mod} into Eq.~\eqref{eq:OCA}, we find that $\chi$ is again composed of a single, now dispersive pole, with the following momentum-dependent energy and residue:
\begin{equation}
    \chi_{\mathrm{mod},i}(q,\omega) = -q^2\left[\frac{R_i(q)}{(\omega+i\eta)^2-\Omega^2(q)}+\frac{C_i}{\epsilon_{b,i}(q)} \right]
\end{equation}

\begin{equation}\label{eq:model_en_disp}
    \Omega_i^2(q) =\Omega_{0,i}^2 + \frac{\vcoul^{\ZD}(q)q^2  R_{0,i}}{\epsilon_{b,i}(q)}, \quad 
    R_i(q) = \frac{R_{0,i}}{\epsilon_{b,i}^2(q)},
\end{equation}
where $\epsilon_{b,i}(q) = 1-v^{\mathrm{2D}}(q)q^2C_{i}$ is the effective dielectric function of the non-absorptive background.
As this model is agnostic with respect to the nature of the quasiparticle involved, it applies both to phonons and excitons, elucidating the analogies between different treatments in the literature~\cite{Thibault_2017,qiu_21}.
From Eq.~\eqref{eq:model_en_disp}, we have $\Omega_i(\mathbf{q}\to 0)=\Omega_{0,i}$ and $R_i(|\mathbf{q}\to 0|) = R_{0,i}$.

In Fig.~\ref{fig:disp_hBN} we also compare the model with $C_i=0$ (without background) and $C_i \neq 0$ (with background)  to emphasize the role of the screening term.
By Taylor-expanding the pole dispersion in Eq.~\eqref{eq:model_en_disp}, we find that $\Omega_i$ disperses linearly with linear coefficient $d\Omega/dq = \pi R_{0,i}/\Omega_{0,i}$.
This corresponds to setting $C_i=0$ in Eq.~\eqref{eq:model_en_disp}.
The linear dispersion coefficient only depends on its residue in $\sigma(\omega)$.
This information nicely explains how optical properties may be linked to the EEL dispersion properties.
The inclusion of a screening non-absorptive term is then mandatory to properly describe the saturation features of the quasi-particle dispersions, elucidating also its nature. The same holds for the intensity decay, properly described by the model conductivity.
We note the linear coefficient of the exciton quasi-particle dispersion is higher than the one which would have been obtained from ab-initio finite difference procedures. We note an approximately $10\times$ denser grid would have been necessary, which requires a prohibitive computational cost for a quantity obtainable from just one $q=0$ calculation with a relatively coarse grid.
Thus, OCA reveals itself also as a useful computational tool to evaluate low momentum-transfer properties between the numerical points of an ab-initio calculation.

\begin{figure}[t] 
    \centering
\includegraphics[width=\textwidth/2, keepaspectratio]{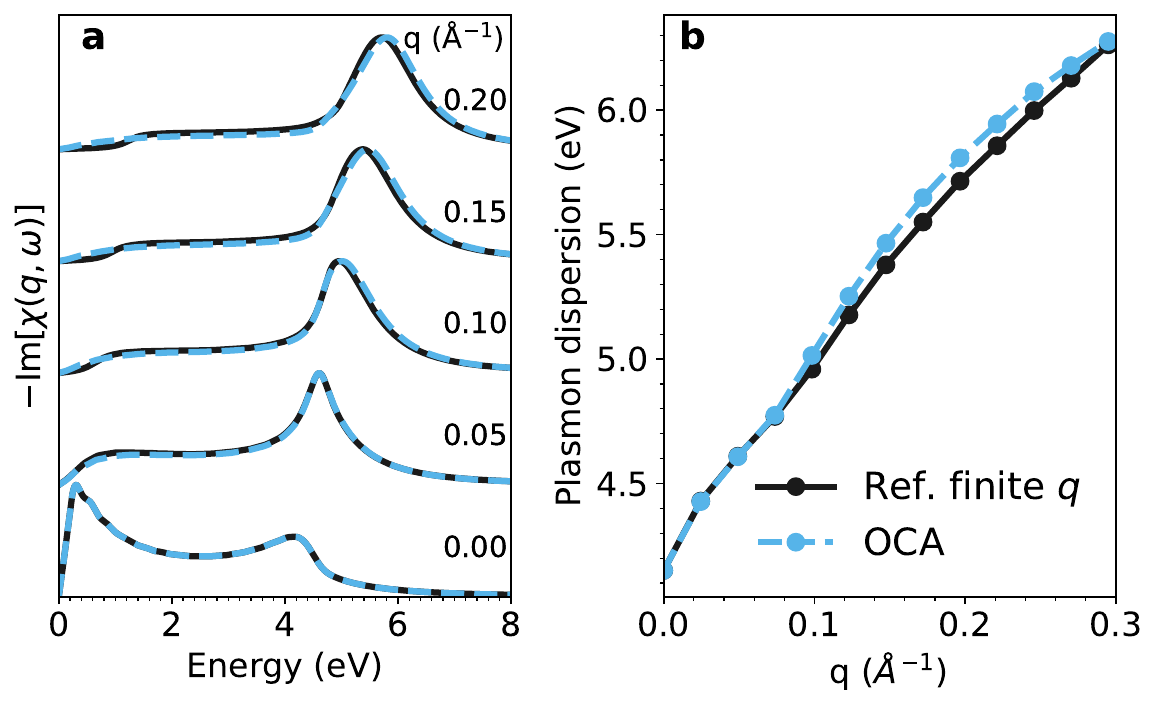}
    \caption{(a-d): loss functions of graphene at different momentum transfer obtained with ab-initio GW-BSE calculations (black) and with the optical conductivity approximation (blue).
    (e): dispersion of the $\pi$-plasmon of graphene obtained with the same methods as in panels (a-d).}
    \label{fig:spectra_gra}
\end{figure}

In a previous work~\cite{guandalini_2023}, we showed both theoretically and experimentally that the graphene $\pi$-plasmon disperses linearly with momentum transfer, and that excitonic effects were needed to describe such a behavior.
In Fig.~\ref{fig:spectra_gra}, we show the ability of OCA to reproduce the linear dispersion of the graphene $\pi$-plasmon, indicating it originates from the dispersion of the 2D Coulomb interactions.
Thus, the only excitonic effects needed to explain the physical properties are those related to the optical conductivity at $q=0$, discussed for example also in Ref.~\cite{Yang2009}.
We notice that OCA is not able to capture the onset dispersion, predicting a null gap at all momentum transfers.
The excitation onset in graphene is determined by the (approximately) linear behavior of the Dirac cone. It is thus a purely band effect, very different in nature with respect to the dispersion of the $\pi$-plasmon which involves transitions around the $M$ point of the Brillouin zone.
This is mainly evident at $q=0.15, 0.20$ \AA$^{-1}$ in Fig.~\ref{fig:spectra_gra}.
In this context, OCA accuracy provides useful information about the physical origin of the spectrum dispersion.

In order to study the effect of an increasing  number of layers on the exciton properties, and to extend the analysis already done for LO/TO phonons~\cite{Thibault_2017}, in the End Matter we show the optical conductivity and $A$ exciton dispersions of bi-layer hBN.

A finite-q Taylor expansion of $\sigma$, although improving the agreement at very small $q$, can violate the positivity of $\mathrm{Re}[\sigma]$ at intermediate momenta; details are given in the Supplemental Material.

\textit{Conclusions---} We have introduced the optical conductivity approximation (OCA), a unified framework for describing low-momentum longitudinal excitations in 2D materials. Applied to monolayer and bilayer hBN and graphene, OCA reproduces the dispersions of LO phonons, bright excitons, and the graphene $\pi$ plasmon, showing that their linear low-\(q\) dispersion is primarily determined by the 2D Coulomb interaction rather than by the momentum dependence of the optical conductivity.

Beyond its theoretical implications, the proposed method offers significant practical advantages. Computationally, since OCA requires only the optical conductivity $\sigma(\omega)$ calculated at $\mathbf{q}=0$, it allows for obtaining accurate low-momentum EELS spectra without the need of demanding calculations on dense momentum grids. 
Experimentally, by inverting the OCA framework, we are able to extract from low-momentum EEL spectra the full complex optical conductivity. This approach, already applied for mono- and few-layer hBN~\cite{Jinhua_2025}, allows one to access optical properties over a broad energy range, circumventing common limitations of standard optical measurements, such as the challenge of reaching high energies and the requirement for large, high-purity crystalline samples to achieve a sufficient signal-to-noise ratio.

Finally, we point out that the proposed scheme equally holds for semiconducting and semimetal systems, as shown explicitly, and eventually for metallic systems, where the $\sqrt{q}$ dispersion of the metallic plasmon is given exactly by the same macroscopic term of the Coulomb potential.

\textit{Acknowledgment---} A.G., P.B. and F.M. acknowledge the funding from European Research Council (ERC) under the European Union’s Horizon 2020 research and innovation program (MORE-TEM ERC-SYN project, Grant agreement No. 951215).
A.F. and D.V. acknowledge partial support by the European Union through the MaX -- MAterials design at the eXascale --
Centre of Excellence (Grant agreement No. 101093374), cofunded by Italian MIMIT, and by Italian MUR
through PRIN 2022“TUNES” (Grant no. 2022NXLTYN).

\renewcommand{\emph}{\textit}
\bibliography{bibliography}
\appendix
\section{End matter}
\begin{figure*}[t] 
    \centering
    \includegraphics[width = 1.0\textwidth, keepaspectratio]{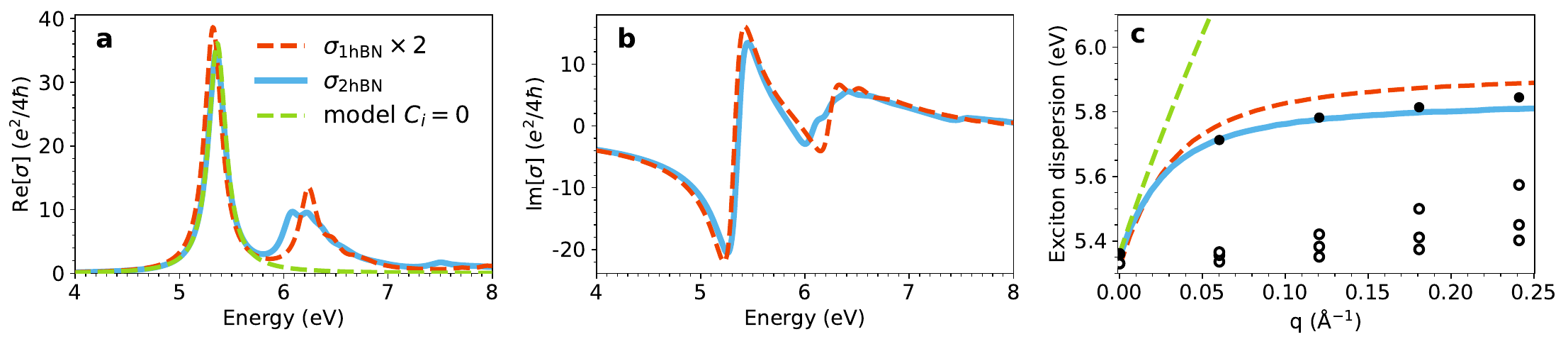}
    \caption{Real (a) and imaginary (b) parts of the optical conductivity of bilayer hBN. (c): bright exciton ($A$) branch dispersion. We show in black the ab-initio dispersions (the bright/longitudinal excitation with filled circles and dark/transverse excitations with empty circles), while the other results are obtained with the optical conductivity approximation. In blue, obtained with the full $\sigma(\omega)$ obtained from ab-initio calculations. In purple, results obtained with $\sigma_{2\mathrm{hBN}}(\omega) = 2\sigma_{1\mathrm{hBN}}(\omega)$, thus with the hypothesis that the hBN layers interact only through the macroscopic Coulomb interaction. In red, results obtained with a model optical conductivity as in Eq.~\eqref{eq:sigma_mod}, with $C_{i}=0$.}
    \label{fig:spectra_2hBN}
\end{figure*}

\textit{Bi-layer hBN---}
In Fig.~\ref{fig:spectra_2hBN} we show the optical conductivity (a,b) and $A$ exciton dispersions (c) of bi-layer hBN.
Due to Davydov's splitting~\cite{Paleari_2018}, the $A$ degenerate branch of monolayer hBN splits into four excitonic states, three dark and one bright.
The $A$ bright peak is located at the same position of monolayer hBN,  due to the decrease of both the GW gap and the exciton binding energy by a similar amount, caused by the additional screening of the second layer.
The $B$ peak acquires a more complex structure due to the same Davydov's splitting applied to each  exciton resonance composing it.
Interestingly, OCA performs equally well for mono and bi-layer hBN, enforcing its validity to reproduce 2D low-momenta dispersion properties of a variety of system classes.

In a picture where the two layers interact only through the macroscopic Coulomb interaction, we have $\sigma_{\mathrm{NhBN}}(\w) \approx N \sigma_{\mathrm{hBN}}(\w)$, in analogy to what happens for phonons~\cite{Thibault_2017}.
This picture works fine for the $A$ exciton branch. 
As the conductivity intensity doubles, also the linear dispersion coefficient is doubled, as predicted by the model conductivity in Eq.~\eqref{eq:model_en_disp}.
Such a behavior, known for phonons~\cite{Thibault_2017}, is consistent with recent experimental measurements~\cite{Peiyi_2026}.
This trend emphasizes that the similar behavior shared between all poles of $\chi$ whose dispersion is dominated by the 2D macroscopic Coulomb interaction.
Instead, the $N\sigma$ model fails to describe the $B$ excitonic branch, due to its more complex nature of being composed by several excitons.

\end{document}